%% file: sample-authordraft.tex
\documentclass[manuscript]{acmart}
\AtBeginDocument{%
  \providecommand\BibTeX{{%
    \normalfont B\kern-0.5em{\scshape i\kern-0.25em b}\kern-0.8em\TeX}}}

\setcopyright{acmcopyright}
\copyrightyear{2018}
\acmYear{2018}
\acmDOI{XXXXXXX.XXXXXXX}

\acmConference[HT '23]{Make sure to enter the correct
  conference title from your rights confirmation emai}{September 04--08,
  2023}{Rome, IT}
\acmBooktitle{ACM HT'23: ACM Hypertext,
 September 04--08, 2023, ACM HT'23, Zoom} 
\acmPrice{15.00}
\acmISBN{978-1-4503-XXXX-X/18/06}

\graphicspath{{figures/}{pictures/}{images/}{./}}
\usepackage[english]{babel} 
\usepackage{amsmath}
\usepackage{textgreek}
\usepackage[scientific-notation=true]{siunitx}
\usepackage{threeparttablex}
\usepackage{multirow}
\usepackage[most]{tcolorbox}
\usepackage{caption}
\usepackage{subcaption}
\usepackage{footmisc}
\usepackage{hyperref}

\tcbset{textmarker/.style={%
        enhanced, parbox=false, boxrule=0mm, boxsep=0mm, arc=0mm, outer arc=0mm, left=4mm, right=2mm, top=7pt, bottom=7pt, toptitle=1mm, bottomtitle=1mm, oversize}}

\definecolor{yellows}{rgb}
{0.98,0.88,0.63}
\definecolor{yellowy}{rgb}
{0.98,0.95,0.89}
\definecolor{greens}{rgb}
{0.6,0.69,0.43}
\definecolor{greeny}{rgb}
{0.94,0.95,0.91}
\newtcolorbox{resultbox}{textmarker, borderline west={6pt}{0pt}{yellows}, colback=yellowy!10!yellowy}
\newtcolorbox{hypobox}{textmarker, borderline west={6pt}{0pt}{greens}, colback=greeny!10!greeny}

\newcommand{\pb}[1]{\vspace{0.75ex}\noindent{\bf \em #1}\hspace*{.3em}}

\begin{document}

\title{Ghost Booking as a New Philanthropy Channel: A Case Study on Ukraine-Russia Conflict}

\author{Fachrina Dewi Puspitasari}
\affiliation{%
  \institution{Korea Advance Institute of Science and Technology}
  \country{Republic of Korea}
}
\email{fachrina.puspitasari@kaist.ac.kr}

\author{Gareth Tyson}
\affiliation{%
  \institution{Hong Kong University of Science and Technology}
  \country{Hong Kong}
}
\email{gtyson@ust.hk}

\author{Ehsan-Ul Haq}
\affiliation{%
  \institution{Hong Kong University of Science and Technology}
  \country{Hong Kong}
}
\email{euhaq@connect.ust.hk}

\author{Pan Hui}
\affiliation{%
  \institution{Hong Kong University of Science and Technology	}
  \country{Hong Kong}
}
\email{panhui@ust.hk}

\author{Lik-Hang Lee}
\authornote{Lik-Hang Lee is the corresponding author: lik-hang.lee@polyu.edu.hk. Fachrina Dewi Puspitasari was a research student in Lik-Hang Lee's lab during the study of the paper.}
\affiliation{%
  \institution{The Hong Kong Polytechnic University}
  \country{Hong Kong}
}
\email{lik-hang.lee@polyu.edu.hk}






\renewcommand{\shortauthors}{Puspitasari, et al.}

\begin{abstract}
The term \textit{ghost booking} has recently emerged as a new way to conduct humanitarian acts during the conflict between Russia and Ukraine in 2022. The phenomenon describes the events where netizens donate to Ukrainian citizens through \textit{no-show} bookings on the Airbnb platform. Impressively, the social fundraising act that used to be organized on donation-based crowdfunding platforms is shifted into a sharing economy platform market and thus gained more visibility.
Although the donation purpose is clear, the motivation of donors in selecting a property to book remains concealed. Thus, our study aims to explore peer-to-peer donation behavior on a platform that was originally intended for economic exchanges, and further identifies which platform attributes effectively drive donation behaviors.
We collect over 200K guest reviews from 16K Airbnb property listings in Ukraine by employing two collection methods (screen scraping and HTML parsing). Then, we distinguish ghost bookings among guest reviews.
Our analysis uncovers the relationship between ghost booking behavior and the platform attributes, and pinpoints several attributes that influence ghost booking. Our findings highlight that donors incline to credible properties explicitly featured with humanitarian needs, i.e., the hosts in penury. 
\end{abstract}

\ccsdesc[500]{Networks~Online social networks}
\ccsdesc[300]{Human-centered computing~User models}

\keywords{Airbnb, Sharing Economy, Donation, Humanitarian acts, Social Computing}



\maketitle

\section{Introduction}
\input{section/Introduction}

\section{Background \& Related Work}~\label{sec:Literature}
\input{section/Literature}

\section{Data Collection Methodology}~\label{sec:Data}
\input{section/Data}

\section{Regression Methodology} ~\label{sec:Analysis}
\input{section/regression}

\section{Understanding Ghost Booking From Platform Attribute}~\label{sec:Result}
\input{section/Result}

\section{Conclusion and Discussion}~\label{sec:Conclusion}
\input{section/Conclusion}

\bibliographystyle{ACM-Reference-Format}
\bibliography{sample-base}

\clearpage

\setcounter{secnumdepth}{0}
\appendix
\section{Suppemental Data}~\label{sec:Supplemental}
\input{section/Appendix}

\end{document}

%% file: section/Introduction.tex
The term ``ghost booking'' has emerged as a new way to
conduct humanitarian acts during the conflict between Russia and
Ukraine in 2022.
The phenomenon started shortly after the introduction of martial law by the Ukrainian government~\cite{fremer_2022}, flooding the Airbnb platform with bookings for Ukrainian properties by guests who (anecdotally) had no intention of staying~\cite{kelleher_2022}.
Ghost booking served as a mechanism for people to donate money to support those living in Ukraine.

In the past, Airbnb has participated in humanitarian aid efforts using top-down approaches, organized by their leadership.
For example, it provided emergency shelters for the evacuees of the Colorado flooding in 2013 and Hurricane Sandy in 2012 through coordination with hosts near the disaster area~\cite{pipetone_2012}. 
In contrast, ghost booking takes a bottom-up approach, driven by individual users. During this ghost booking drive, Airbnb hosts are using their property listing to solicit donations, as shown in Figure~\ref{fig:appeal-flowchart}(a). 
This represents a rather new form of humanitarian support, which hitherto has remained unstudied.

This paper aims to explore which factors impact the likelihood of receiving a ghost booking donation on Airbnb.
We borrow from the principle of signaling theory.
The social theory tries to explain communication behavior between a signaler and receiver who experience information asymmetry~\cite{spence1973}.
Accordingly, the information is classified as either \textit{signals} (host-generated information such as booking fee, property type, host tenure--how long a host has been renting, and listing title and description) or \textit{feedback} (guest-generated information such as booking reviews, host qualification, and ratings). 
We dissect the relationship between these attributes and the likelihood of receiving a ghost booking (that we estimate using the number of ghost booking reviews).

We discover several influential factors that correlate with the probability of receiving a ghost booking. These include the host qualification (e.g., Superhost), the inclusion of solicitation for donations in the title or description, property type, count of normal booking reviews, Superhost tenure, and booking fee.
Our study contributes to the understanding of platform donations, in the context of Airbnb characterized by a bottom-up approach.

After depicting the background and related work (Section \ref{sec:Literature}), the paper begins with the methodologies of collecting data regarding Airbnb listing in Ukraine and labeling ghost booking with Latent Dirichlet Allocation (LDA) (Section~\ref{sec:Data}). 
Section~\ref{sec:Analysis} presents the analysis by understanding the characteristics of ghost booking reviews given the platform attributes, and further provides statistical and semantic analysis. 
Accordingly, Section~\ref{sec:Result} highlights the key findings.
Last, Section~\ref{sec:Conclusion} summarizes our findings as well as the potential drawbacks of utilizing ghost booking as a philanthropy channel.

%% file: section/Literature.tex
We first review the definition of ghost booking and the characteristics of booking transactions on Airbnb. We also outline the platform attributes from the theoretical perspective.

\pb{The Understanding of Ghost Booking.}
The utilization of Airbnb as a medium for social fundraising is a relatively new concept. 
Ghost booking is the term used to describe a person who books an accommodation from an online platform but does not physically use it. 
The movement was first initiated by a Twitter user named Quentin Quarantino\footnote{\url{https://twitter.com/quentquarantino/status/1499441114738212871?cxt=HHwWjoCy2fvvis8pAAAA}} at the beginning of March 2022.
Shortly after, the tweet went viral, and netizens began to donate to Ukrainian Airbnb hosts via ghost bookings.

\pb{The Nature of Transactions in Airbnb.}
Normally, Airbnb is utilized for the purpose of booking accommodation for a short stay. Former studies have explored the characteristics of Airbnb.
For example, the booking fee and former guest reviews are found to be important factors for Airbnb users obtaining a booking~\cite{varma2016airbnb}. Moreover, other studies discovered that the number of guest reviews is considered more meaningful for the guest than the property rating~\cite{lee2015analysis}. The same study also confirmed that newer hosts have more popularity~\cite{lee2015analysis}. However, other researchers found that guest reviews may contain bias due to manipulation (system selection to eliminate scams) or natural selection (rule of \textit{big number} or the already-low guest expectation)~\cite{fradkin2015bias}. In regards to listing characteristics, a former study found a negative relationship between the booking fee and the booking attraction. This study also discovered that guests often prefer private accommodation (i.e., entire property) rather than a shared space because it enables them to have more freedom~\cite{zhang2019listening}. The same characteristics were also found in another study, which found that guests are more attracted to properties that provide detailed information~\cite{yao2019standing}.

\pb{Signaling Information on Airbnb.}~\label{sub:Signal}
One of the widely used theories in explaining the communication patterns between two parties is the \textit{signaling theory}.
Although it was first developed in the field of economics by Michael Spence in 1973~\cite{spence1973}, the main idea still revolves around the need to minimize information asymmetry~\cite{connelly2011signaling}. 

The main construct of this theory consists of two actors (the signaler and receiver) and two pieces of information (the signal and feedback). Here, one cycle of communication is divided into four periods~\cite{connelly2011signaling}. In the first period, signalers prepare the information that they want to convey. Next, signalers send it to the receiver, contained in a signal. During the third period, receivers accept the signals, interpret them, and prepare the response. Finally, receivers deliver them back to the signaler, bundled as feedback.

Similar principles 
have been applied in the study of Airbnb~\cite{yao2019standing}. Thus, our work aligns with the signaling theory
to characterize the attributes of the Airbnb platform with regard to the ghost booking phenomenon. We aggregate the attributes into \textit{signals} if contents originated from the hosts themselves. Examples of such attributes are the listing description, pictures, booking fee, list of amenities, host identity, and types of rental accommodation. Conversely, we group the Airbnb features into \textit{feedback}, which covers information from anyone but the hosts themselves. These attributes may be in the form of former guest reviews, ratings, and host qualifications. 
Although, in practice, these features are still shared by the hosts to Airbnb users, theoretically, they are part of the feedback. 
Host qualification feature (superhost vs. regular host) is released by the Airbnb system, but we still consider them  in the feedback group as the system refers mainly to the data of former guests' reviews and evaluations when assessing the host performance. Note that the Superhost badge (Figure~\ref{fig:appeal-flowchart}(a)) can be achieved within a year of having an average rating of at least 4.8/5, at least ten bookings, a cancellation rate below 1\%, and a response rate within 24 hours for at least 90\% of new inquiries~\cite{airbnb_superhost}.

\pb{Prior studies on Airbnb.}~\label{sub:airbnb}
Since Airbnb is the most representative and successful example of the sharing economy, significant research efforts can be seen in the past decade, in terms of the economic incentives and two-sided market designs~\cite{www-2017-market-design, www-2019-market-mechanism}. In addition, peer-to-peer interaction among hosts and guests becomes a vital source of user-generated data. The data analytics of such data can act as novel socioeconomic indices, such as the cues of understanding longstanding residential communities featured with cultural diversity and authenticity~\cite{CSCW-NOV22}, the 
beneficiary of Airbnb in a local community~\cite{WWW2016-BenefitsAirbnb}, as well as a real-time indicator of tracking neighborhood changes, i.e., a socioeconomic index~\cite{CSCW-APR21}. 

Other researchers focus on the relationship between people and sharing economy services. Giovanni \textit{et al.} conducted a 10-year analysis to explore the guests' reviews from six cities located in the United States, the United Kingdom, and Australia. 
The analysis highlights social interactions in Airbnb. 
The social interaction between the hosts and guests can deteriorate due to uncertainty, as the two parties can only interact with each other through self-disclosure profile pages and guest reviews. Such self-disclosure signals become the primary conveyor of trust~\cite{CSCW2017-FEB}. It is also worthwhile mentioning that potential guests can spend hours and days examining the description of the listing~\cite{WWW19-AirbnbMetric}.
Similarly, Lampinen and Cheshire~\cite{CHI-16-Airbnb} conducted a qualitative study to examine the potential of introducing a third party to reduce uncertainty between Airbnb hosts and guests, and hence their financial assurance in the emerging social interaction. In addition, Nata \textit{et al.} proposed an experimental interface that enables users to invest in trust-driven tokens, and hence constructs a behavioral framework for building trustworthy information in sharing economy platforms~\cite{WWW20-Designing-Trust}. 

Our work studies the features of Airbnb's Ukrainian listings as confounding factors for the number of ghost booking reviews. In addition to confirming the findings from previously studied features such as host qualification, our work highlights other features, such as donation solicitation. In summary, our analysis offers a unique and the \textit{first} case study on humanitarian support through Airbnb (sharing economies) based on the Russia-Ukraine crisis, as compared to the well-known yet organized crowd-sourced donation through diversified platforms~\cite{charitable-tech-cscw08}. The traditional platforms include DonorsChoose\footnote{\url{https://www.donorschoose.org}} and CrowdFunder,\footnote{\url{https://www.crowdfunder.co.uk}} while the emerging platforms primarily refer to live streaming interaction like Twitch~\cite{www-22-twitch-donation}, Facebook, and YouTube~\cite{www20-donation-livestreaming}. 
The most relevant works refer to the design of mechanisms, information conveyors, and interfaces, which drive the willingness or persuasiveness of donation on these platforms~\cite{donate-are-you-willing-MUC20,how-to-ask-a-donation-UMAP22, inequality-donation-www19, cscw-altruism-donate-2020, cscw21-multiplatform}.

%% file: section/Data.tex
\subsection{Airbnb Dataset}~\label{sec:Dataset}
\begin{figure}[t]
    \centering
    \includegraphics[width=\linewidth]{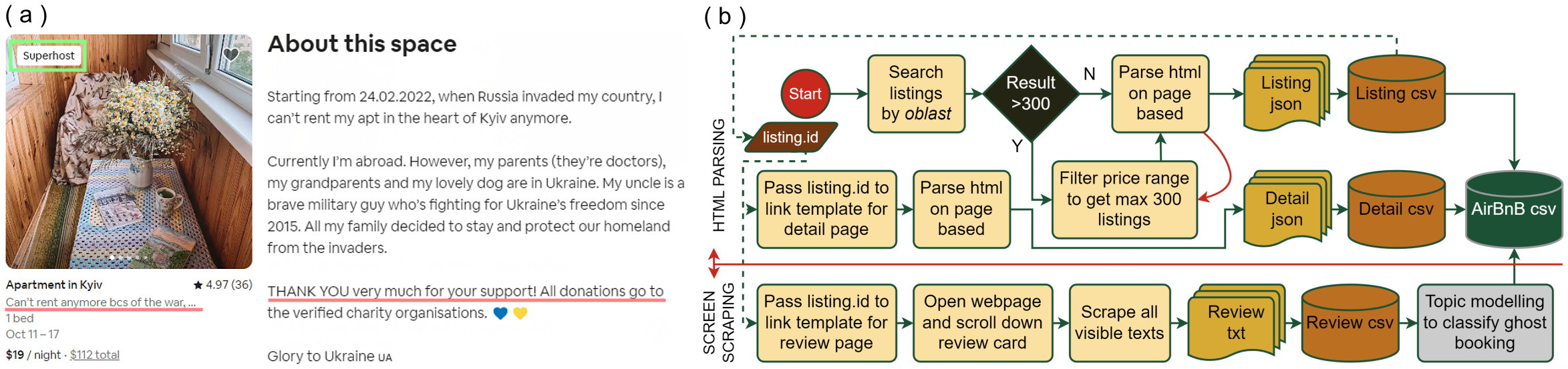}
    \caption{An example of an Airbnb property listing highlighting the call for support and donations (a). Red underlines show the donation solicitation (\textit{Signal}) in the title and description. A users' 
    Superhost badge (\textit{Feedback}) is also visible in green box. This information is collected by employing the data collection flowchart in (b).}
    \Description{Figure 1}
    \label{fig:appeal-flowchart}
\end{figure}

To examine ghost booking behavior in Airbnb, we collect publicly available data from the website starting from September 2\textsuperscript{nd} to 20\textsuperscript{th}, 2022. 
Figure~\ref{fig:appeal-flowchart}(b) describes the procedures of data collection, while Table~\ref{tab:feature} (Supplemental Data)
contains a summary of the data collected.

\pb{Accommodation Listings.}
First, we gather the list of available accommodations from the search page. Our search term is the name of 24 Ukrainian \textit{oblasts}\footnote{This is an administrative division in Ukraine, \url{https://en.wikipedia.org/wiki/Oblast}} as destinations. 
To maximize our search result, we set the dates to ``flexible'' for a one-week stay (\textit{a week} period is the most reasonable choice for a short stay compared to \textit{only a weekend} and \textit{a month}). 
Since Airbnb limits the search to fifteen pages (twenty listings per page), there is a maximum of 300 listings displayed per search. Thus, to retrieve all accommodation data, we filter the search by price range.
For each oblast, we manipulate the range of booking fees to make each search result into 300 listings.
This dataset covers general information about the property, such as the listing ID, available date, location, listing title, property type, booking fee, average rating, review count, picture hyperlink, and host qualification.

\pb{Accommodation Details.}
Using the listing IDs from the above search page dataset, we then visit the accommodation pages to extract all public information. This covers the property description, reviews, list of amenities, and host profile. 
    
\pb{Guest Reviews.}
Finally, we gather all property reviews by scrolling through the review cards on each accommodation page. Each review provides information related to the guest name, posting date, review comments, and host responses. We collect all the reviews, regardless of their posting date. Later, we use the count of reviews as a proxy for the number of bookings. 

The final dataset consists of 16,330 unique Ukrainian properties with a total of 209,466 guest reviews, ranging from October 2011 to September 2022.

\subsection{Labeling Ghost Reviews}~\label{sec:Classification}
We next label each review as a ghost vs. non-ghost booking indicator. 
The former refers to a guest review, which is likely from a guest who intentionally did not stay on the property; while the latter signifies otherwise. While non-ghost booking reviews may be present at any time period, we posit that ghost booking reviews only started following the Ukraine-Russia conflict.
For simplicity, we refer to non-ghost bookings as ``\emph{normal bookings''}.
We next describe how we categorize reviews into these two groups.

\pb{Topic Extraction.}
Prior to the analysis, we pre-process the Airbnb guest reviews by removing the account handles ('\textit{Response from [host name]}') and boilerplate text.
Other unrelated sentence elements, such as tabs, new lines, punctuation, symbols, and emojis, are also removed. Last, reviews with empty content are excluded from the dataset.

We then perform LDA using the \textit{textmineR} and \textit{LDAvis} libraries.
First, we tokenize and stem the pre-processed reviews.
Subsequently, we create a document term matrix using the combination of unigrams or bigrams of the stem. 
The topic model is constructed by selecting the number of topics (17) using the optimum coherence score.
After the appropriate model has been constructed, the topics are assigned to each review. We provide examples of five topics in Table~\ref{tab:topic} in the 
Supplemental Data. There are topics (topic number 3, 15, and 16) related to the ongoing crisis with keywords such as \textit{support} and \textit{love}. Topic numbers 8 and 11 show a more generic Airbnb-specific discourse highlighting the comfort and location of the listed properties.

\pb{Temporal Analysis of Topics.}
To identify the emergence of ghost bookings, we use the topics assigned to each review to identify a significant change in review content.
Our conjecture is that these topic changes might be indicative of the switch from `normal' bookings to ghost bookings. 

\begin{figure}[t]
     \centering
     \includegraphics[width=\linewidth]{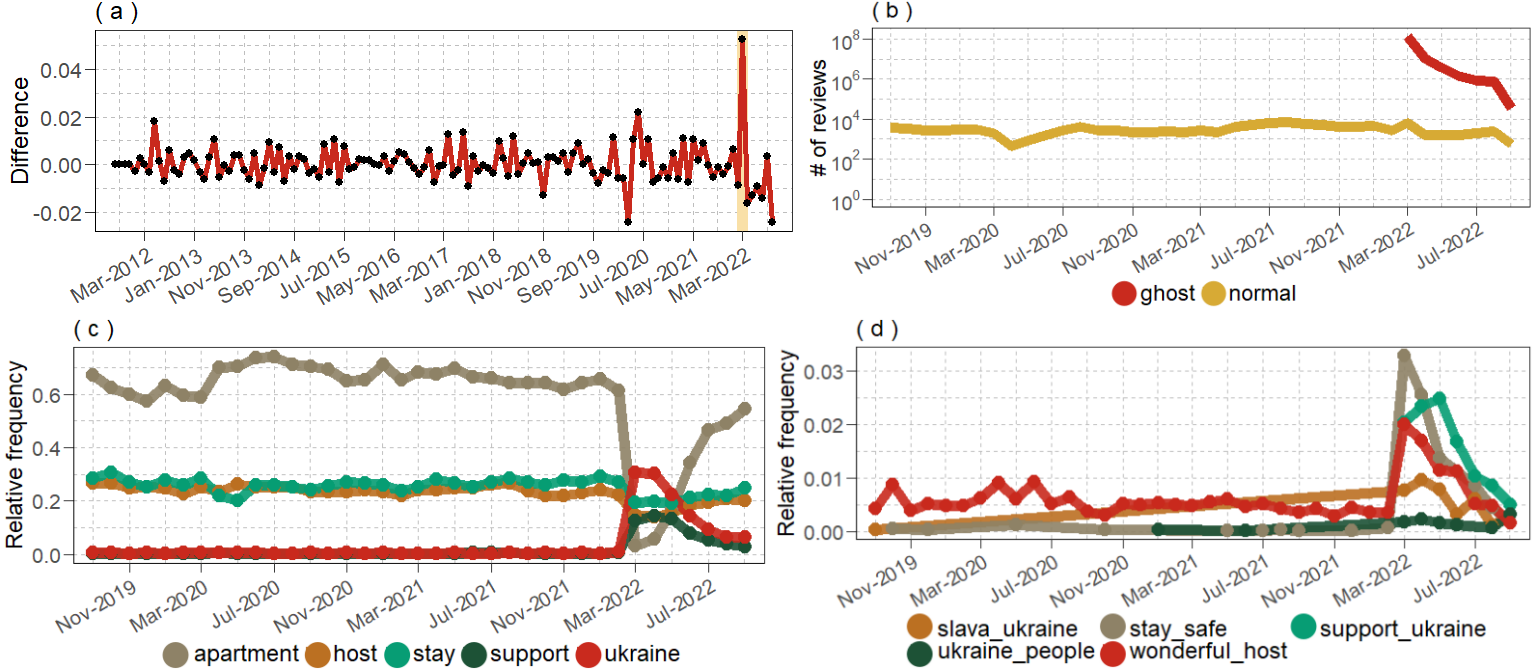}
     \caption{Data profile; in (a), inter-topic distance discrepancy measured between month-\textit{n} and month-\textit{n-1}. Largest deviation in topics occurs between February to March 2022.; in (b), volume of ghost and normal booking reviews. Monthly average ghost booking reviews count sextuples normal booking ones.; in (c and d), top-5 unigrams and bigrams of all guest reviews since March 2022, respectively. All terms, except \textit{apartment}, \textit{host}, and \textit{stay}, experience >100\% inflation or sudden appearance in March 2022.}
     \Description{data-collected}
     \label{fig:data-collected}
 \end{figure}

We divide the dataset into months (note that it is impossible to subdivide further, as Airbnb only gives monthly timestamps for reviews).
We then compute the average inter-topic distances between each contiguous time segments. Figure~\ref{fig:data-collected}(a) shows the result plotted on a time series graph. As distance measures the degree of difference among topics, the y-axis in this graph represents the change between the distance of month \textit{n} and month \textit{n-1}. There is a clear spike between February and March 2022. The average Euclidean distance of reviews in March increases a remarkably 650\% from February. This shows that the topics discussed within these periods changed dramatically.

\pb{Ghost Review Classification.}
Inspired by the above, we argue that the text topics can serve as a mechanism to identify ghost reviews.
To classify a topic as indicative of a ghost booking, we extract the terms that experience an abrupt spike since the start of the conflict.
We first extract the top-5 frequent unigram and bigrams from all guest reviews starting from March 2022.  
Subsequently, we analyze their temporal relative frequency over time. Our technique is to compute their frequency of occurrence in each time segment and divide by the total number of reviews in the respective segment. Figures~\ref{fig:data-collected}(c) and \ref{fig:data-collected}(d) depict the relative usage frequency of the top-5 unigrams and bigrams, respectively. We only show the data from November 2019 to September 2022 for better readability. 
Starting from March 2022, the terms indicative of ghost reviews are those that suddenly appear or increase $>$100\% in relative usage compared to the previous time segment.
Specifically, these terms are \textit{support}, \textit{ukraine},  \textit{support\_ukraine}, \textit{stay\_safe}, \textit{wonderful\_host}, \textit{ukraine\_people}, and \textit{slava\_ukraine}.
Table~\ref{tab:unibigram} (Supplemental Data)
presents the statistics for this selection.
Accordingly, we refer to these terms to categorize the LDA topics from March 2022 into ghost booking indicative by manually scrutinizing the top-20 terms of each topic.

Next, we label each review as ghost vs.\ normal. 
We further verify the accuracy by randomly checking against ten reviews in each time segment starting from March 2022.
Through this, we obtain a total of 31,816 ghost booking reviews and 114,629 normal booking reviews (Figure~\ref{fig:data-collected}(b)). In other words, the total ghost reviews accumulated within seven months is almost 30\% of the total normal booking reviews for twelve years.

%% file: section/regression.tex
We use the property features as described in Section~\ref{sec:Literature} to quantify their effect in relevance to the number of ghost booking reviews (as a proxy of the number of bookings). To this end, we train a regression model with the number of ghost-booking reviews as the dependent variable. The choice of regression analysis is based on its interpretability of the effect for independent variables.
We use negative-binomial regression due to the non-normal distribution of the data~\cite{blasco‐moreno_pérez‐casany_puig_morante_castells_2019}.

\pb{Model building.} 
We follow an incremental approach and build 5 regression models in total. We first start with three models:
Model 1 involves variables categorized as feedback: count of normal booking reviews, and host qualification. 
Model 2 includes variables grouped as signals: booking fee, property type, hosting tenure, listing title, and listing description. Finally, we incorporate all independent variables into Model 3, to test the influence of both signals and feedback on ghost booking likelihood. 

We then pick the best-fitting model from these three models and add two additional variables (host qualification and listing title) interaction terms to create Models 4 and 5, respectively. 
Our grouping is based on the signaling theory discussed in Section~\ref{sec:Literature}.
Our goal is to investigate whether donors are more influenced by the information published by a host on the Airbnb platform (signals), or by the facts affirmed by the previous guests (feedback). Table~\ref{tab:model} (Supplemental Data)
lists the variables included in our models.

\pb{Selection of Most-Fitting Model.} We use the Akaike information criterion (AIC) metric to select the best-fitting model from the first iteration with three models. Note that AIC measures the relative quality of three models, and the lowest metric indicates the most-fitted model as it indicates that the model has the least information loss relative to the true model. Model 3 has the lowest AIC metric (47,782.23) and is therefore chosen as the best model.
Additionally, this model also has the highest goodness-of-fit value (log-likelihood ratio test value of 3,191.29), indicating that it greatly deviates from the null model. Table~\ref{tab:model} (Supplemental Data)
shows the result for all three models; the significant variables are those with (\textalpha) below 0.05. 

We observe from Model 3 that all main independent variables, except host tenure, fall in the 95\% confidence band, indicating their significance to the volatility of ghost booking review count. Additionally, we also notice that the regression coefficients for host qualification and listing title are the two highest among others, indicating that both cause the greatest change in the number of ghost booking reviews. Hence, due to their large coefficient value (host qualification \textbeta\textsubscript{2} = 1.390 and listing title \textbeta\textsubscript{6} = 1.132), we suspect that there exists a multicolinearity between them and other independent variables, which may influence the effect of the main variables in predicting ghost booking.

According to the above observation, Model 3 may have limitations. One of these is that it only considers the change in the number of ghost booking reviews caused by each platform attributes, given everything else is constant. Such an assumption may weaken the external validity of the model because guests conduct a \textit{trade-off} assessment when choosing which accommodation to book. Thus, relying only on the main variables may overlook the effect of co-linearity between these variables. To overcome this, we extend Model 3. 
We put the interaction effects of property advert title and host qualification with the main variables into Models 4 and 5, respectively. The use of two separate models will better highlight their effect on the number of ghost-booking reviews likelihood. A generalized formula for the final model can be written as:

\begin{displaymath}
\begin{aligned}
        Ln(GH)=\: &\alpha\, + \beta_{k}v_k + \beta_{n}v_k*i_j + \epsilon
\end{aligned}
\end{displaymath}

\noindent
where $v_k \in V$ and $V$ is the set of all independent variables. $i_j \in I$ is a set of interaction variables. $\beta_k$ and $\beta_n$ are the coefficients for each term with $K = $ number of all variables and $N = $number of all possible interaction combinations.

%% file: section/Result.tex
We next explore which features are most determinant in predicting the number of ghost-booking reviews a property receives.
To achieve this, recall, we rely on Model 5, based on the lowest AIC measure (47,468.02), to explain our findings. 
In the context of Airbnb, we define the seven attributes as either \textit{signal} and \textit{feedback}. Title of Donation Solicitation, Property Type, Description of Donation Solicitation, Superhost Tenure, and Booking Fee refer to Signal, as the Airbnb hosts offer these descriptive yet one-way attributes to pitch potential donors. On the other hand, Host Qualification and the Number of Normal Booking Reviews are regarded as feedback because this information is based on user-generated data derived from the interaction between Airbnb hosts and guests.

\subsection{Host Qualification}~\label{sec:Superhost}
\begin{figure}[t]
    \centering
    \includegraphics[width=\linewidth]{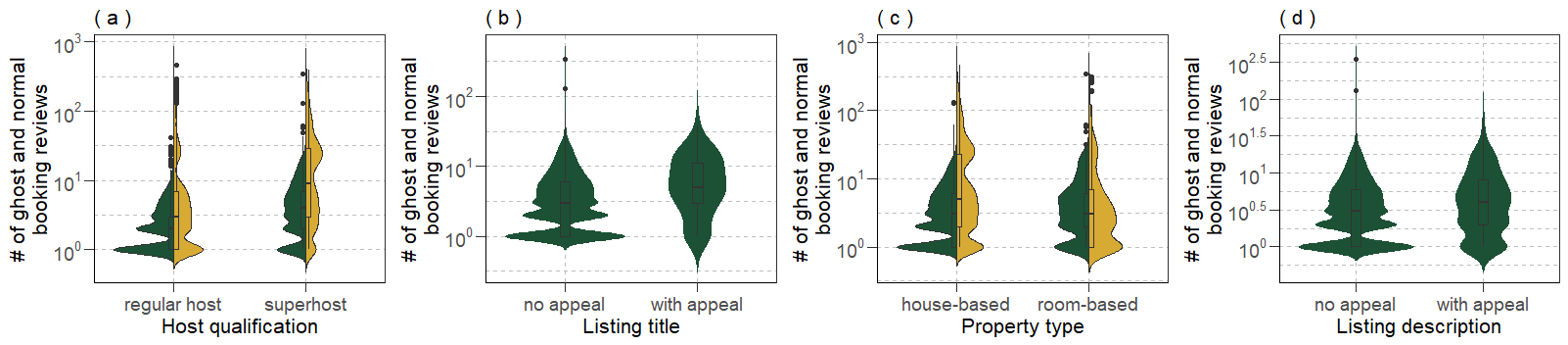}
    \caption{Violin plots of ghost (green) and normal (yellow) booking reviews for host qualification (a), donation solicitation in title (b), property type (c), and donation solicitation in description (d). The patterns of the number of ghost booking reviews, on average, is having higher occurrence on properties that are: room-based, Superhost-rented, and solicit donation in the title and the description.}
    \Description{categorical}
    \label{fig:categorical}
\end{figure}

Host qualification (feedback) emerges as the most important attribute based on our regression result (\textbeta\textsubscript{2} = 1.5375).
Thus, Figure~\ref{fig:categorical}(a) plots the distribution of the number of ghost (green) and normal (yellow) booking reviews for both Superhost and regular host qualifications. 
It shows that the median and interquartile range of the number of ghost booking reviews on Superhost properties are higher than regular host ones.
We also observe a similar pattern for the distribution of the number of normal booking reviews. 
However, since both the median and interquartile range of the number of normal booking reviews are higher than ghost ones, we suspect the number of ghost and normal booking reviews have a different distribution.
We next conduct the Chi-square goodness of fit test and find that both the number of ghost vs. normal booking reviews on Superhost and regular host properties similarly produce small p-value (<0.05 with \textchi\textsuperscript{2} of 40,854 vs. 39,399).
This demonstrates that both types of booking reviews are different yet are likely to occur more on Superhost properties.
This also reinforces our regression results, i.e., we see that Superhosts get 6.35x more ghost booking reviews than regular hosts.

This pattern may indicate that donors are most concerned about donating only to trustworthy individuals.  As it is unlikely the host and donor know each other, the Superhost status is one of the few mechanisms to establish trust. 
Further, responsiveness (as required for the Superhosts status) can serve as a key criterion for trustworthiness~\cite{jones2012trustworthiness}. 
To support this, we compare the relative occurrence of the terms that are indicative of host responsiveness written on Superhost vs. regular host properties.
A former study on the exploration of Airbnb guest reviews content lists \textit{question}, \textit{quick}, \textit{respond}, \textit{answer}, and \textit{prompt} as terms indicative of host's responses~\cite{zhang2019s}. 
Referring to this study, we find that the relative frequency for the top-3 terms (\textit{quick}, \textit{respond}, and \textit{prompt}) are all higher on Superhost properties than the normal host ones (1.35\% vs. 1.32\%; 2.19\% vs. 2.00\%; and 0.27\% vs. 0.26\% respectively).
Further details are listed in Table~\ref{tab:semantic} (Supplemental Data).

\subsection{Donation Solicitation in Title}~\label{sec:Title}
Each property has an associated title (signal), e.g., ``Spacious One Bedroom Flat".
We conjecture that some titles may actively solicit contributions (e.g., ``donations welcome").
Indeed, titles that contain donation solicitation with words like (\textit{donate}, \textit{support}, \textit{help}, and \textit{charity}) are the second most important features in attracting ghost booking reviews (\textbeta\textsubscript{6} = 0.9648). 
Figure~\ref{fig:categorical}(b) shows the distribution of the number of ghost booking reviews (green) for listing titles with and without solicitation.
Here, the median for the former is higher than the latter.
The distribution of the number of ghost booking reviews for the title with solicitation is concentrated around the median (4) and the third quartile (8.5), while the distribution for the counter-part converges around the lower quartiles (median = 0 and third quartile = 2).

This demonstrates that ghost booking is likely to occur in property listings that request donations in their titles.
Overall, we see that properties that solicit donations in their title receive 2.82x more ghost booking reviews compared to others.

The above suggests that the title may help donors to select receivers as they can easily filter hosts seeking financial assistance during the Ukraine-Russia conflict.
This observation is in line with former studies on online feedback systems, where the provision of descriptive information helps the buyer's purchase decision~\cite{siering2018explaining}.
By explicitly requesting donations in the title, a host may signal to the donors that the property is accepting aid (also known as ``needs signaling''~\cite{connelly2011signaling}).

\subsection{Property Type}~\label{sec:Room}
Property type (signal), room vs. entire house, is the third most important feature in our regression analysis (\textbeta\textsubscript{4} = 0.4160). This shows that room-based properties may receive about 1x more ghost-booking reviews than house-based ones.
Figure~\ref{fig:categorical}(c) presents the distribution of the number of ghost (green) and normal (yellow) booking reviews for room-based vs. house-based properties. 
We notice that the distributions are rather different.
The first quartile of the number of ghost booking reviews on house-based properties tends to be lower (high concentration of 1 review) than room-based ones.
The average number of ghost booking reviews for house-based properties is 1.67, which can be compared to 2.22 for room-based properties.
On the other hand, the interquartile range for the number of normal booking reviews is higher in house-based properties than the room-based ones.
We validate this observation using the Chi-square goodness of fit test for both the number of ghost vs. normal booking reviews on room-based and house-based properties. As a result, 
a small p-value (<0.05 with \textchi\textsuperscript{2}\textsubscript{room-based} = 15,392 vs. \textchi\textsuperscript{2}\textsubscript{house-based} = 82,811) indicates that the difference does exist.

The contrasting pattern of the number of ghost booking reviews on room-based vs. house-based properties can be caused by the former offering lower booking fees than the latter. Our data shows that the average booking fee for room-based (\$50.63 USD) is lower than house-based (\$80.57 USD).
Naturally, a lower fee may attract more users to book the property than the counterpart of expensive booking fees. 
Besides, since the room-based listings refer to a shared accommodation (e.g., a private room with a shared toilet), donors possibly are not concerned over the privacy issue~\cite{zhang2019listening} as they do not plan to stay there.

\subsection{Donation Solicitation in Description}~\label{sec:Description}
Donation solicitation in the description of the property listing (signal) is the next most powerful confounding factor on the number of ghost-booking reviews (\textbeta\textsubscript{7} = 0.2367). 
This solicitation attracts 0.46x more ghost booking compared to other listings that do not solicit.
Figure~\ref{fig:categorical}(d) illustrates the distribution of the number of ghost booking reviews (green) on property descriptions with and without solicitation.
The plot shows a similar pattern as observed in Section~\ref{sec:Title}.
On average, the number of ghost booking reviews is higher for those with solicitation than those without.
However, since the medians of the number of ghost booking reviews over listing descriptions with vs. without donation solicitations are not greatly different (1 vs. 0), the influence of donation solicitation in listing descriptions towards ghost booking is potentially weak.

\subsection{Number of Normal Booking Reviews}
\label{sec:Review}
\begin{figure}[tb]
    \centering
    \includegraphics[width=\linewidth]{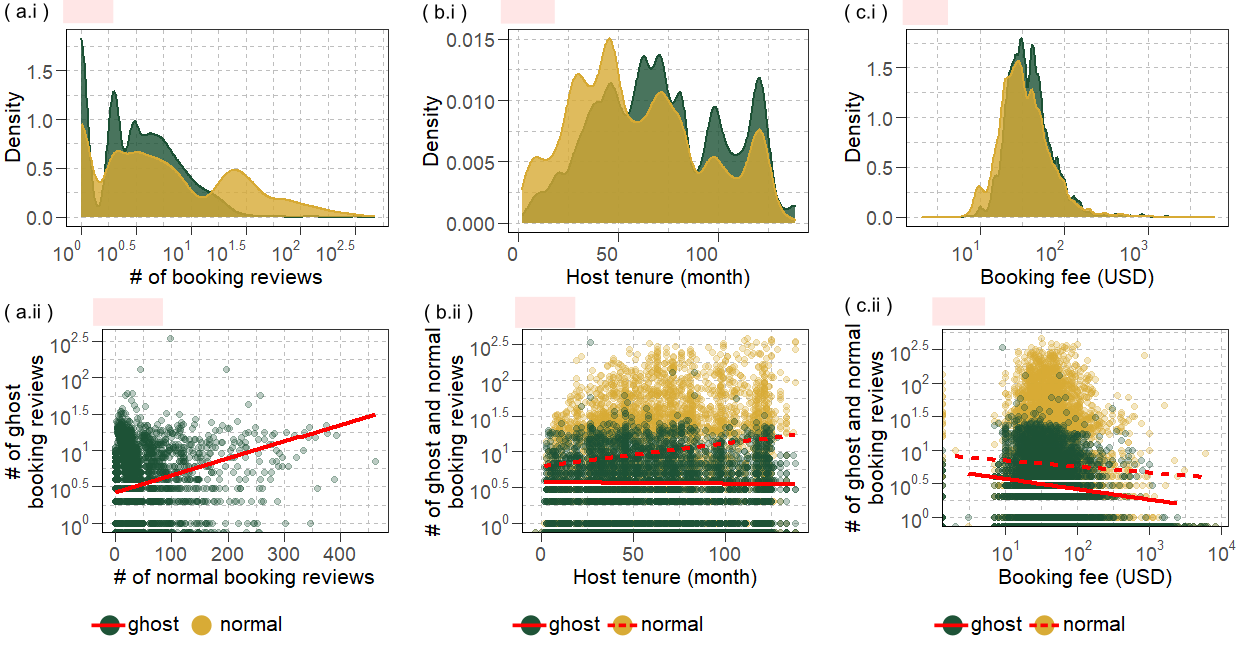}
    \caption{Probability density function (a.i) and scatter plot (a.ii) of ghost booking reviews for the number of normal booking reviews. The continuous line represents the mean of ghost booking reviews count. The number of ghost booking reviews grows following the increase in normal booking reviews count. Probability density function (b.i) and scatter plot (b.ii) of ghost or normal booking reviews for Superhost tenure in month. Means of ghost and normal booking review counts on Superhost tenure are illustrated by continuous and dashed lines, respectively. The magnitude of growth of the number of normal booking reviews is higher than the degree of decline of the number of ghost booking reviews. Probability density function (c.i) and scatter plot (c.ii) of ghost or normal booking reviews for booking fee in USD. Means of ghost and normal booking reviews count are illustrated by continuous and dashed lines, respectively.  The number of ghost and normal booking reviews shrink following the growth of booking fee.}
    \Description{continuous}
    \label{fig:review-tenure-fee}
\end{figure}

Contrary to the normal booking behavior in Airbnb~\cite{varma2016airbnb},
the count of normal booking reviews (feedback) has little predictive power for the number of ghost bookings (\textbeta\textsubscript{1} = 0.0241).
Figure~\ref{fig:review-tenure-fee}(a.i) further illustrates the density of distributions for the number of ghost (green) and normal (yellow) booking reviews, in which both distributions are exhibit a different number of peaks; the ghost booking reviews have sharper peeks as compared to the normal booking review.
The small p-value (<0.05, U = 72,405,036) from the Mann-Whitney test further supports this discrepancy.
Furthermore, Figure~\ref{fig:review-tenure-fee}(a.ii) presents the relationship between the number of ghost (green) and normal booking reviews.
From this scatter plot, we notice a slightly increasing trend in the number of ghost booking reviews following the increase in the number of normal booking reviews.
We further evaluate the correlation of both types of the number of booking reviews in Figure~\ref{fig:correlation}(a).
Here, we observe that the number of ghost booking reviews positively correlates (0.26), although slightly weak, with the number of normal booking reviews.
This also supports our regression result where we observe that the mean count of the number of ghost booking reviews only increases by 2.44\% subjected to 1 increase in the number of normal booking reviews.

We infer the weak correlation pattern between both numbers of booking reviews as potentially caused by the difficulty to derive further information solely from the numerical measures of normal booking reviews.
A prior study in Airbnb also found that while having more reviews may indicate the popularity of a listing, the count of reviews is less influential in increasing the booking odds than their actual content~\cite{yao2019standing}.

\subsection{Superhost Tenure}
\label{sec:Tenure}
Our regression shows that the tenure of a Superhost (signal) may impact the number of ghost reviews (\textbeta\textsubscript{17} = -0.0036).
This is slightly different from the normal booking pattern where guests tend to book hosts (regardless of qualification) with shorter tenure~\cite{lee2015analysis}.

To further investigate this, Figure~\ref{fig:review-tenure-fee}(b.i) shows the distribution of the Superhost tenure, for the number of ghost (green) and normal (yellow) booking reviews.
We observe that both distributions are different; this is confirmed through a Mann-Whitney U test, which results in a small p-value (<0.05, U = 1,107,668,184).
We see that the average Superhost tenure of the ghost review properties is 60.66 months vs. 59.25 months for the normal reviews.

Figure~\ref{fig:review-tenure-fee}(b.ii) shows the relationship between both types of booking reviews count and the Superhost tenure. 
We observe that the number of ghost booking reviews decreases slightly after the increase in host tenure.
We further observe the correlation between Superhost tenure and the number of ghost booking reviews in Figure~\ref{fig:correlation}(b), which shows that they are very weakly correlated to each other (-0.03).

\subsection{Booking Fee}~\label{sec:Fee}
Surprisingly, the booking fee (signal) has a very weak influence on the likelihood of receiving ghost booking reviews (\textbeta\textsubscript{3} = -0.0009).
To further investigate this, Figure~\ref{fig:review-tenure-fee}(c.i) shows the distribution of the booking fee for the number of ghost (green) and normal (yellow) booking reviews.
We notice that both distributions are different, as confirmed by the small p-value (<0.05, U = 2,164,319,869) of a Mann-Whitney U test. 
Additionally, Figure~\ref{fig:review-tenure-fee}(c.ii) shows the relationship between the number of ghost (green) and normal (yellow) booking reviews with booking fee in USD.
Although both types of booking reviews decrease following the increase in booking fee, the magnitude of the decline in the number of ghost booking reviews is larger than the normal ones.

\begin{figure}[t]
    \centering
    \includegraphics[width=\linewidth]{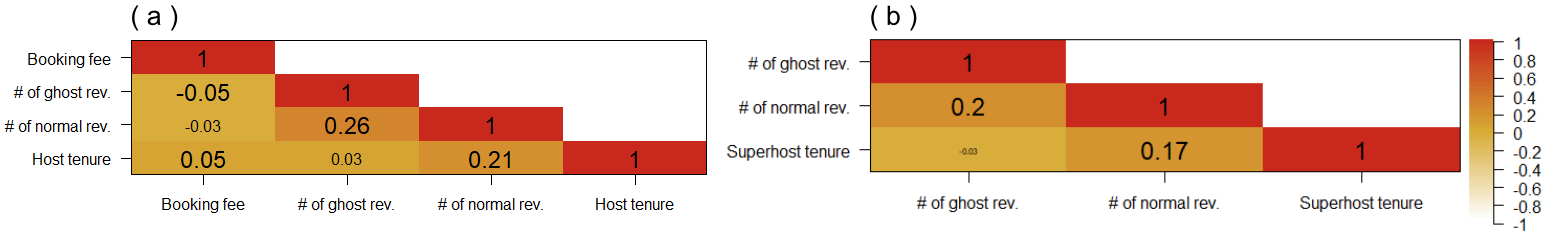}
    \caption{Correlation matrix; in (a), among platform features with continuous values. The number of ghost booking reviews correlates very weakly with host tenure and booking fee and more strongly with the count of normal booking reviews.; in (b), the Superhost tenure and the number of ghost and normal booking reviews. Correlation values from (a) are moderated.}
    \Description{distribution}
    \label{fig:correlation}
\end{figure}

Figure~\ref{fig:correlation}(a) reinforces this relationship, showing that the correlation score (although very weak) of the booking fee with the number of ghost booking reviews (-0.05) is slightly higher than the normal ones (-0.03).
This reduction is also supported in our regression result that observes the number of ghost booking reviews undergoes a slight decrease (0.09\% of mean count) with \$1 USD increase in the booking fee.

We conjecture that these differences are due to the different expectations between ghost and normal booking.
While real guests who plan to visit the properties look forward to the experience, donors may have fewer concerns over such matters.
Thus, naturally, donors may not consider the trade-off between booking fee and stay experience.
Hence, they might simply seek a booking fee that approximately matches how much they wish to donate.

%% file: section/Conclusion.tex
This paper has explored the phenomenon of ghost booking on Airbnb, as a mechanism to donate to people in Ukraine. 
We have examined which factors impact the likelihood of receiving ghost booking reviews (as a proxy for the number of bookings).

\pb{Findings.}
As the platform attributes are classified into \textit{feedback} (information originated from guests as receiver) and \textit{signal} (details generated by hosts as signaler), our study discovers that host qualification (feedback), either regular or Superhost, emerges as the most vital platform feature that donors consider when booking. 
It is important to note that Superhost gets 6.35x more reviews from ghost booking donors than regular hosts. 
Additionally, the property advert title (signal) also appears to be vital in influencing donors' decisions to book. 
We found that hosts who solicit donation in their listing titles get 2.82x more ghost booking reviews than those who do not.

Our observations on Airbnb data imply that both feedback and signal are equally important in affecting the decision to donate, conforming to the previous studies on their effect on donation~\cite{ho2021influence}. Host qualification is an algorithmically aggregated term that provides a holistic performance review of the host on the Airbnb platform. However, this measure is indicated together with other signals, i.e., property features. For instance, the super host badge appears on the top corner of the property photo (as shown in Figure~\ref{fig:appeal-flowchart}(a).  Naturally, more information leads to a more minor degree of information asymmetry, and thus reinforcing the quality of the host's signals~\cite{courtney2017resolving}. Nevertheless, feedback, albeit only two feedback types are studied in this paper, is potentially more potent than signal, as seen from the different numbers of ghost booking reviews that both can attract.

\pb{Potential Pitfalls of Ghost Booking.}
Trustworthiness is always the critical issue of donation between the donors and receivers, and thus impacts the robustness of the donation mechanism. Remarkably, the issue is well recognized by recent studies~\cite{jones2012trustworthiness, CSCW2017-FEB, WWW20-Designing-Trust, donate-are-you-willing-MUC20}. We acknowledge that both the signals and feedback act as the basis for informing the potential donors and hence establishing the donors' perception of trustworthiness on a particular host. Although our work attempted to parse the signals and feedback, but such cues may manipulate the donors' perception of the trustworthiness on a particular hosts. Donors who donate through ghost booking may still need to be aware of whether their donation is right-on-target for the following reasons.

\textit{First}, when taking the Superhost badge for granted, donors may be misled only by trustworthiness without knowing the host's financial ability. Our result in Section~\ref{sec:Result} shows that there are a number of Superhost who have been hosting the property for a long time. For such hosts, donors may become slightly hesitant to donate (decrease 0.36\% from the mean count) because they may consider them financially sufficient. Furthermore, the host might rent multiple properties in Ukraine or anywhere else in the world.

\textit{Second}, the distribution of aid may become uneven since naturally, not all Ukrainian households rent on the Airbnb platform. Referring to our statistics (Section~\ref{sec:Data}), given there are only 16,330 Ukrainian listings, assume no hosts renting multiple properties, and take the size of each household for 2.46\footnote{\url{https://en.wikipedia.org/wiki/List_of_countries_by_number_of_households}\label{population}}, ghost booking may only be able to help at maximum (all properties receive ghost booking) 40,172 citizens out of 41,723,998\footref{population} or less than 1\% of total populations.

\pb{Limitations and Future Work.}
We highlight the limitations of our work and present the future line of research. 
We do not include guest-specific features in our analysis; as such, it is limited by our data collection approach. 
Future work with guest information may provide a holistic approach from both donors and receivers.  
Accordingly, we are keen to perform a longitudinal study on both guests and hosts during the ghost booking campaign and its effects on bookings.

For future work, we will further elicit relevant user-perceived attributes prevalent on traditional (top-down) and crowdsourced fundraising platforms, to further generalize our findings. Accordingly, we will leverage such attributes to consider the potential mechanism of a fundraiser certification system for donation activities during other new crises. In other words, certified hosts can facilitate humanitarian acts during warfare or natural disasters, especially when the purposes of such sharing economy platforms are initially not designated for humanitarian responses. On the other hand, the attributes of signals and feedback can be further considered as the socioeconomic indices for humanitarian response, e.g., hosts' trustworthiness and donors' willingness to donation.

%% file: section/Appendix.tex
\begin{table}[!htb]
    \caption{Airbnb features extracted. Features from listings are collected from the search page, while the ones from details and reviews are collected from detail page.}
    \label{tab:feature}
    \begin{threeparttable}
        \begin{tabular}{lll}
            \toprule
                \textbf{Feature} & \textbf{Description} & \textbf{Source} \\
            \midrule
                Listing ID & Unique listing identity number & Listings \\
                Date & Available week for booking & Listings \\
                Location & Coordinate position of property & Listings \\
                Title & Listing title & Listings \\
                Property & Type of accommodation offered & Listings \\
                Price & Booking fee per night & Listings \\
                Rating & Average rating of property & Listings \\
                Review & Volume of reviews & Listings \\
                Link & Hyperlink for listing pictures & Listings \\
                Badge & Host qualification & Listings \\
                Description & Listing description & Details \\
                Amenity & Count of amenities & Details \\
                Host tenure & Host joining date & Details \\
                Ghost* & Volume of ghost booking reviews & Reviews \\
                Normal* & Volume of normal booking reviews & Reviews \\
            \bottomrule
        \end{tabular}
        \begin{tablenotes}
            \small
            \item *Features added after LDA classification. 
        \end{tablenotes}
    \end{threeparttable}
\end{table}

\begin{table}[!htb]
  \caption{Description of main variables used in the regression models. Model 1 includes \textit{feedback} variables, Model 2 includes \textit{signal} variables, Model 3 includes all main variables. Model 4 and 5 add the set of interaction variables between listing title and host qualification, respectively, to the set of main variables.}
  \label{tab:variable}
  \begin{tabular}{ll}
    \toprule
    \textbf{Variable} & \textbf{Description} \\
    \midrule
    \multicolumn{2}{l}{\textbf{Response}}\\
    \hspace*{2mm}Ghost booking review & \# of ghost booking reviews\textsuperscript{a} \\
    \multicolumn{2}{l}{\textbf{Predictor}}\\
    \multicolumn{2}{l}{\hspace*{1mm}\textit{\textit{Feedback}}}\\
    \hspace*{2mm}Normal booking review & \# of normal booking reviews\textsuperscript{a} \\
    \hspace*{2mm}Host qualification & Superhost or regular host status\textsuperscript{b} \\
    \multicolumn{2}{l}{\hspace*{1mm}\textit{\textit{Signal}}}\\
    \hspace*{2mm}Property type & Type of accommodation offered\textsuperscript{b} \\
    \hspace*{2mm}Listing title & Title with donation terms\textsuperscript{b} \\
    \hspace*{2mm}Listing description & Description with donation terms\textsuperscript{b} \\
    \hspace*{2mm}Booking fee & Cost of property rent per night\textsuperscript{c} \\
    \hspace*{2mm}Host tenure & Period of hosting in month\textsuperscript{c} \\
    \bottomrule
  \end{tabular}
  \begin{tablenotes}
        \footnotesize
        \item \textsuperscript{a,b,c}Indicate discrete, categorical, and continuous data respectively. 
      \end{tablenotes}
\end{table}

\begin{table}[b]
  \caption{Sample of LDA topic from May 2022. Topic no. 3, 15, and 16 are indicative to ghost booking.}
  \label{tab:topic}
  \begin{tabular}{ll}
    \toprule
    \textbf{Topic no.} & \textbf{Top-5 terms} \\
    \midrule
    3 & support, ukraine, people, support\_ukraine, book \\
    8 & comfort, clean, house, location, bed \\
    11 & center, location, nice, city, clean \\
    15 & stay, hope, safe, family, stay\_safe \\
    16 & ukraine, love, stand, send, glory \\
    \bottomrule
  \end{tabular}
\end{table}

\begin{table}[b]
    \caption{Statistics of relative frequency (RF) for top-5 unigrams and bigrams indicative to ghost booking.  All terms, except \textit{apartment}, \textit{host}, and \textit{stay}, experience >100\% inflation or sudden appearance in March 2022.}
    \label{tab:unibigram}
    \begin{threeparttable}
        \begin{tabular}{lcc}
            \toprule
                \textbf{N-grams} & \textbf{*Contiguous RF} & \textbf{**STD before March '22} \\
            \midrule
                apartment & -95\% & 4\% \\
                host & -34\% & 4\% \\
                stay & -28\% & 6\% \\
                support & 1813\% & 38\% \\
                ukraine & 3145\% & 54\% \\
                slava\_ukraine & 2831\% & - \\
                stay\_safe & 4432\% & 73\% \\
                wonderful\_host & 452\% & 25\% \\
                support\_ukraine & Sudden & - \\
                ukraine\_people & 730\% & - \\
            \bottomrule
        \end{tabular}
        \begin{tablenotes}
            \small
            \item *Difference between RFs in March and February 2022 divided by RF in February 2022. 
            \item **Standard deviation for proportion of difference among contiguous RFs before March 2022. 
        \end{tablenotes}
    \end{threeparttable}
\end{table}

\begin{table}[b]
  \caption{Summary statistics for all variables. Zero median for the number of ghost booking reviews indicates over-dispersion.}
  \label{tab:summary}
  \begin{tabular}{lccccc}
    \toprule
        \textbf{Variable} & \textbf{mean} & \textbf{med} & \textbf{sd} & \textbf{min} & \textbf{max}\\
    \midrule
        Ghost booking rev. & 1.74 & 0.00 & 4.78 & 0.00 & 345.00 \\
        Normal booking rev. & 10.67 & 1.00 & 29.42 & 0.00 & 462.00 \\
        Host qualification & 0.43 & 0.00 & 0.49 & 0.00 & 1.00 \\
        Booking fee & 76.89 & 39.00 & 211.07 & 2.00 & 8181.00 \\
        Property type & 0.13 & 0.00 & 0.34 & 0.00 & 1.00 \\
        Host tenure & 55.39 & 47.00 & 32.63 & 0.00 & 140.00 \\
        Listing title & 0.005 & 0.00 & 0.07 & 0.00 & 1.00 \\
        Listing description & 0.05 & 0.00 & 0.22 & 0.00 & 1.00 \\
    \bottomrule
  \end{tabular}
\end{table}

\begin{table}[b]
  \caption{Summary of semantic analysis as indicator of \textit{responsiveness}. Higher relative frequency for all terms on Superhost indicates that higher responsiveness.}
  \label{tab:semantic}
  \begin{tabular}{llllll}
    \toprule
    \textbf{Term} & \textbf{Category} & \textbf{\# of reviews} & \textbf{Freq.} & \textbf{Relative freq.} \\
    \midrule
    Quick & Superhost & 22,590 & 306 & 1.35\% \\
    Respond & & & 496 & 2.19\% \\
    Prompt & & & 60 & 0.27\% \\
    Quick & Regular host & 5,840 & 77 & 1.32\% \\
    Respond & & & 117 & 2.00\% \\
    Prompt & & & 15 & 0.26\% \\
    \bottomrule
  \end{tabular}
\end{table}

\begin{table*}[b]
  \caption{Negative binomial count parameter for ghost booking likelihood. Model 5 with the lowest AIC score presents 7 significant attributes affecting the likelihood of ghost booking reviews count.}
  \label{tab:model}
  \begin{threeparttable}
      \begin{tabular}{lcccccccc}
        \toprule
            \textbf{Variable} & \textbf{Coefficient} & \textbf{Model 1} & \textbf{Model 2} & \textbf{Model 3} & \textbf{Model 4} & \textbf{Model 5} \\
        \midrule
            \underline{\textit{Normal booking review}} & \textbeta\textsubscript{1} & 0.0199*** & & 0.0241*** & 0.0242*** & \textbf{0.0241***} \\
            \underline{\textit{Host qualification}} & \textbeta\textsubscript{2} & 1.4037*** & & 1.3131*** & 1.3116*** & \textbf{1.5375***} \\
            \underline{\textit{Booking fee}} & \textbeta\textsubscript{3} & & -0.0011*** & -0.0009*** & -0.0009*** & \textbf{-0.0009***} \\
            \underline{\textit{Property type}} & \textbeta\textsubscript{4} & & 0.2909*** & 0.4768*** & 0.4788*** & \textbf{0.4160***} \\
            \textit{Host tenure} & \textbeta\textsubscript{5} & & 0.0039*** & -0.0011* & -0.0011* & 0.0009 \\
            \underline{\textit{Listing title}} & \textbeta\textsubscript{6} & & 1.7568*** & 1.1449*** & 0.6157 & \textbf{0.9648*} \\
            \underline{\textit{Listing description}} & \textbeta\textsubscript{7} & & 0.5618*** & 0.2538*** & 0.2410*** & \textbf{0.2367*} \\
            \textit{Listing title x Normal booking review} & \textbeta\textsubscript{8} & & & & 0.0176** & \\
            \textit{Listing title x Host qualification} & \textbeta\textsubscript{9} & & & & 0.5702 & 0.2369 \\
            \textit{Listing title x Booking fee} & \textbeta\textsubscript{10} & & & & 0.0040 & \\
            \textit{Listing title x Property type} & \textbeta\textsubscript{11} & & & & -0.6455 & \\
            \textit{Listing title x Host tenure} & \textbeta\textsubscript{12} & & & & -0.0011 & \\
            \textit{Listing title x Listing description} & \textbeta\textsubscript{13} & & & & 0.7865 & \\
            \textit{Host qualification x Normal booking review} & \textbeta\textsubscript{14} & & & & & 0.0002 \\
            \textit{Host qualification x Booking fee} & \textbeta\textsubscript{15} & & & & & -0.0006 \\
            \textit{Host qualification x Property type} & \textbeta\textsubscript{16} & & & & & 0.1775 \\
            \underline{\textit{Host qualification x Host tenure}} & \textbeta\textsubscript{17} & & & & & \textbf{-0.0036***} \\
            \textit{Host qualification x Listing description} & \textbeta\textsubscript{18} & & & & & 0.0119 \\
        \midrule
            AIC & & 47,694.23 & 50,694.23 & 47,496.00 & 47,509.61 & \textbf{47,468.02} \\
            Log likelihood & & -23,843.11 & -25,312.35 & -23,739.00 & -23,739.79 & -23719.00 \\
            Log likelihood ratio & & 3,269.29 & 330.81 & 3,477.52 & 3,475.93 & \textbf{3,517.52} \\
            p-value$\ddagger$ & & \num{0.0000000000000002} & \num{0.0000000000000002} & \num{0.0000000000000002} & \num{0.0000000000000002} & \num{0.0000000000000002} \\
        \bottomrule
      \end{tabular}
      \begin{tablenotes}
        \small
        \item $\dagger$ * p-value < 0.05, ** p-value < 0.01, *** p-value < 0.001
        \item $\dagger\dagger$Model 5 has the lowest AIC value (most fit among all models) and highest log likelihood ratio test value (most fit compared to null model).
        \item $\dagger\dagger\dagger$Underlined variables have significant relationship with response variable with reference p-value of 0.05.
        \item $\ddagger$ p-value from Chi-square statistics of log likelihood ratio test.
      \end{tablenotes}
    \end{threeparttable}
\end{table*}